\documentclass{aastex}
\usepackage{emulateapj5,graphicx}

\shorttitle{Dust Transport and Pop II Star Formation}
\shortauthors{Venkatesan, Nath, \& Shull}

\begin{document}

\title{The Radiative Transport of Dust in Primordial Galaxies and
Second-Generation Star Formation}
\submitted{Accepted by ApJ, to appear in v. 640, 20 March 2006}

\author{Aparna Venkatesan\altaffilmark{1}, Biman B. Nath\altaffilmark{2,3},   and J. Michael Shull} 
\affil{CASA, Department of Astrophysical and Planetary Sciences, \\
University of Colorado, Boulder, CO 80309-0389}
\altaffiltext{1}{NSF Astronomy and Astrophysics Postdoctoral Fellow}
\altaffiltext{2}{JILA Visiting Fellow, University of Colorado and National
Institute of Standards and Technology}
\altaffiltext{3}{Raman Research Institute, Bangalore - 560080, India}
\email{(aparna, mshull)@casa.colorado.edu, biman@rri.res.in} 
\vspace{0.1in}

\begin{abstract}

We investigate the radiative transport of dust in primordial galaxies in
the presence of the UV radiation field from the first metal-free stars. We
find that dust created in the first supernova (SN) explosions can be driven
through the interior of the SN remnant to accumulate in the SN shells,
where second-generation stars may form from compressed cooling gas. This
scenario requires metal-free stars to form continuously over timescales of
up to 10 Myr, consistent with recent estimates. Silicate and graphite
grains, as well as iron-bearing magnetites, are transported to the shells
for reasonable parameter assumptions, but their relative yields from
primordial SNe is an important factor in the resulting abundance ratios. We
compare the results of segregated grain transport with the current
nucleosynthetic data on extremely metal-poor Galactic halo stars.  Fossil
signatures of this process may already have been detected in those
iron-poor stars with enhanced carbon and silicate elements such as
magnesium, silicon and oxygen. We discuss the implications of our results
for the transition from first- to second-generation star formation in
primordial galaxies, and the role played by the radiative transport of dust
in this process.

\end{abstract}


\section{Introduction}

The radiative transport of dust and grain history in galaxies have been
studied by many authors, generally in the context of our Galaxy (e.g.,
\citealt{drainesalpeter1,dwekscalo,seabshull}). Although early calculations
\citep{pecker72} of this problem indicated that dust transport was
difficult to accomplish, \citet{ferraraetal91} found that it was possible
to transport dust grains by radiation pressure to large distances from the
plane of the disk in a relatively short time ($\sim 10^8$ yr).  In
addition, dust destruction mechanisms are important to consider
\citep{shustov}.  

In general, smaller dust grains are destroyed at a faster rate while they
are being transported, and they typically do not survive.  Grain
destruction in the interstellar medium (ISM) occurs through thermal
sputtering in hot gas as well as non-thermal processes (sputtering,
shattering, vaporization in grain-grain collisions).  The non-thermal
processes occur primarily when the grains are swept up by shock waves
\citep{shull78,jonesetal94}.  In these situations, the original power-law
grain size distribution will be modified by various processes of grain
destruction \citep{seabshull,jonesetal96}. Grains can also be accelerated
through the gas by radiation fields.  However, at a given radiation
pressure, it is more difficult to drive large grains owing to their lower
ratio of area to mass. The final outcome of the radiative transport of dust
therefore depends on the size and composition of the grains as well as the
details of the radiation field.

More recently, numerical simulations in a cosmological context have studied
the radiative ejection of dust grains to the intergalactic medium (IGM)
from primordial halos in the wake of the first generations of supernovae
(SNe) \citep{aguirreetal01,bianchi05}.  Testing the predictions of such
theories of cosmological dust transport has usually been in a macroscopic
context, using upper limits from the cosmic microwave background to
constrain IGM dust distortion \citep{lh97,ferraraetal99} or to
offer alternative explanations to a cosmological constant scenario of the
data on Type Ia SNe at redshifts, $z \la 2$ \citep{aguirre99}.

In this paper, we focus on more local effects of dust transport within the
host galaxy related to second-generation stars that form coevally with or
subsequent to primordial stars. We examine, under the conditions in typical
primordial halos, whether dust can be driven effectively by the radiative
pressure of metal-free stars. If so, can dust grains accumulate in the
cooling shells of first-generation SN remnants (SNRs), becoming
incorporated eventually into new star-forming sites? Recent work by
\citet{mackey03} and \citet{salvaterra04} suggest that such
second-generation star formation is highly likely, forming from metal-poor
or even metal-free fragmenting gas in SN shells in early galaxies. The
latter authors show that instabilities in the SN shell that could lead to
gas fragmentation and to the eventual formation of low-mass second-generation
stars set in at about 1--50 Myr after the SN explosion. This range holds
for the condition that the SN shells do not sweep the baryons out of a
galaxy of virial temperature 10$^4$ K.  A timescale requirement of $\sim$ a
few Myr to $10^8$ yr is entirely consistent with the duration of metal-free
star formation calculated by both semi-analytic \citep{tvs04} and numerical
methods \citep{wadavenk, byh03}. These involve estimates of the timescales
over which metals from the first SNe can enrich the gas in either the host
galaxy or neighboring galaxies.  We motivate this hypothesis by considering
the currently available element abundance ratios of extremely metal-poor
(EMP) stars in the Galactic halo. The elements comprising the dominant dust
compounds from the first SNe, including C, Si, O, and Fe, in addition
provide the most effective cooling channels for primordial star-forming gas
\citep{brommloeb03, santoroshull}.

The paper is organized as follows. In \S\ 2, we present the current
nucleosynthetic data on EMP stars to motivate the consideration of
a dust-transport scenario. In \S\ 3, we describe the formalism and
assumptions of the model used to solve for dust grain transport, the
results of which are presented in \S\ 4.  We discuss implications of our
findings and conclude in \S\ 5.

\section{Nucleosynthetic Data on EMP Stars}

We begin by using the current data on EMP stars relevant for the problem in
this paper. We highlight the trends of carbon, silicon, oxygen, nitrogen,
magnesium, and aluminium as a function of iron abundance below [Fe/H] $\sim
-2.8$. These elements, with the exception of nitrogen, are the most
relevant for dust creation from the metals from the SNe of metal-free stars
\citep{schneideretal04, todini}. We define [Fe/H] as the ratio of the
measured column densities of Fe/H to the solar ratio, (Fe/H)$_\odot$ =
$4.68 \times 10^{-5}$.

The nature of the primordial stellar initial mass function (IMF) is
currently of great interest and debate.  Some recent theoretical studies
indicate that this IMF may have been top-heavy \citep{abeletal00, bromm02},
leading predominantly to stellar masses $\ga$ 100 $M_\odot$ up to a
critical gas metallicity of $Z_{\rm cr} \approx 10^{-4 \pm 1} Z_\odot$
\citep{brommetal01, schneider02}, above which a present-day IMF
occurs. However, other detailed studies of the current data on
reionization, high-$z$ star formation and the metal abundance ratios in the
IGM and EMP stars suggest that the primordial IMF, rather than being biased
towards high masses, may merely lack low-mass stars
\citep{venktruran,tvs04, daigne04, qianwass05}. It is certainly possible
that both IMFs were coeval in the past, given the right combination of
conditions, as pointed out by these papers. In addition, the transition
metallicity $Z_{\rm cr}$ may be significantly higher at low densities and
could vary with metal species \citep{brommloeb03, santoroshull}. These
two IMFs, $\ga$ 100 $M_\odot$ and $\sim$ 10--100 $M_\odot$, represent two
possibilities in the definition of a top-heavy IMF, where the IMF's lower
or upper mass limit is increased. A third option would be to flatten the
slope or alter the shape of the stellar IMF.

Regardless of such IMF issues, we expect the first generations of stars to
form from metal-free gas. Their composition heavily influences their
structure and properties, as they rely predominantly on the p--p chain
initially than on the more efficient CNO cycle for their thermonuclear fuel
source \citep{tumshull}. Consequently, metal-free stars are hotter and emit
significantly harder ionizing radiation relative to their finite-$Z$
counterparts \citep{bkl, tsv, sch02}. This will play an important role for
the dust transport problem here.

Stars of masses $\sim$ 10--100 $M_\odot$ end their lives as the more
familiar Type II SNe, leaving behind neutron stars and black holes, whereas
metal-free stars in the mass range $\sim 140-260 M_\odot$ are thought to
disrupt themselves entirely as pair-instability SNe (PISNe).  We do not
concern ourselves here with the exact type of SN in primordial halos, and
require only that the parent metal-free stellar cluster has effective
luminosities of $\sim 10^6 L_\odot$ \citep{tsv, bkl, sch02}, which is
easily achieved by a few tens of stars in a Salpeter IMF in the 1--100
$M_\odot$ range or a single star of mass $\ga$ 100 $M_\odot$.  For the
purposes of this paper, we require the former case, i.e., a stellar
cluster, so that there is a hard photon source for at least a few to 10 Myr
after the initial SN explosion. We also require that the SN kinetic energy
is at least $10^{51}$ erg, which is true for ``normal'' Type II SNe. We
discuss the case of hypernovae (HNe) below; the SN explosion energies for
HNe and PISNe are thought to lie in the range $10^{51}$--$10^{53}$ erg
\citep{umedanomoto, heger02}. \\ 

\centerline{\epsfxsize=1.0\hsize{\epsfbox{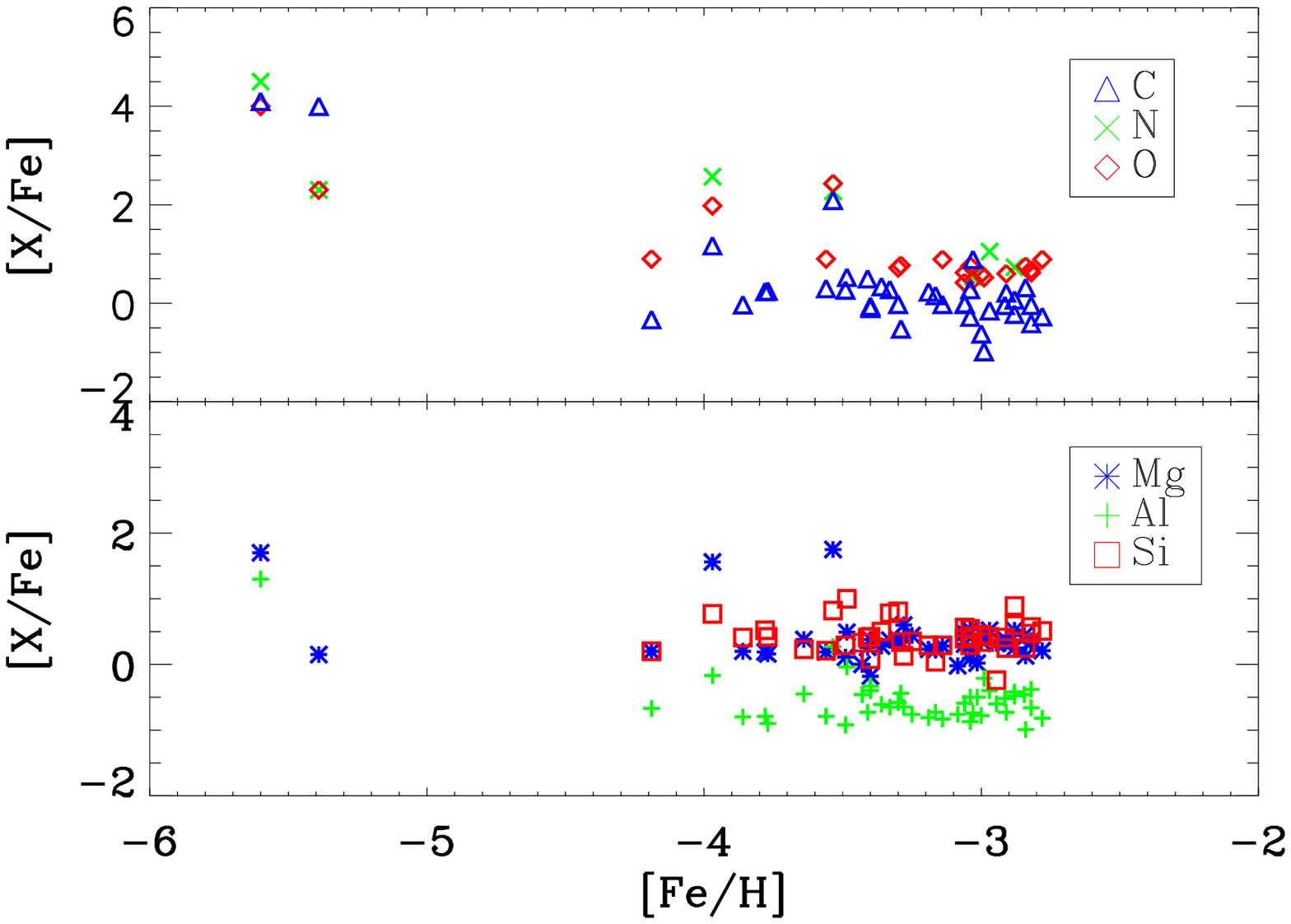}}}

\vspace{0.1in} 
\figcaption{The observed abundances of C, N, O (upper panel)
and Mg, Al and Si (lower panel) relative to Fe as a function of [Fe/H] in
EMP halo stars. See text for discussion.}
\vspace{0.2in} 

In Figure 1, we show the measured EMP stellar abundances of C, N, O, Mg,
Al, and Si relative to Fe (normalized to the solar ratio), as a function of
[Fe/H]. We limit the data to those EMP stars with [Fe/H] $\la -2.8$, to be
consistent with the upper limit to the transition metallicity factoring in
typical errors in the data of order 0.1-0.2 dex. This is also close to the
upper limit in metallicity derived by \citet{salvaterra04} for
second-generation stars forming in the SN-induced scenario. We take the
data of EMP giants from the study by \citet{cayrel}, and of EMP dwarfs and
other stars from \citet{cohenetal04}, \citet{aoki04}, and \citet{tvs04} and
references therein. For the two most iron-poor stars, HE-0107-5240 and the
newly discovered HE-1327-2326, we use the published abundances by
\citet{christlieb04} and \citet{frebel05}.

We note some relevant points of interest in Figure 1. First, although the
two most iron-poor stars have greatly enhanced values of C, N and O, there
is no overall clear trend of [C/Fe] with [Fe/H]. We also observe that those
EMP stars with highly enhanced C (so-called C-enhanced
EMPs)\footnote{Selecting merely on the basis of C-enhancement is unlikely
to narrow down the parent process, as many C-enhanced EMP stars show
selective enhancements of $r$- and $s$-process, as well as other
elements. As many authors have discussed, this great diversity of abundance
ratios may indicate an equal diversity of contributing mechanisms.} also
have strong N overabundances relative to solar, sometimes even exceeding
that of C. Oxygen, on the other hand, is usually enhanced, sometimes up to
10$^2$--$10^4$ times solar at the lowest Fe-metallicities. We note also the
consistent enhancements of Mg and Si of a few to 100 times solar, and the
relatively flat mild underabundance of [Al/Fe] with [Fe/H].

Several models have been proposed to explain the diversity of element
abundances in low-mass EMP stars. Some interpret the abundances as
reflecting the star's gas formation conditions, while others invoke
processes subsequent to its creation. The presence of Mg and heavier
elements would argue for the former class of models in principle.  Amongst
these, the HN model \citep{umedanomoto} is currently the most successful at
reproducing the EMP data, requiring a new class of SNe at primordial
metallicities that have enhanced SN explosion energies of $10^{51} -
10^{53}$ erg and enhanced CNO element generation relative to Fe at low
[Fe/H]. HN models can reproduce most EMP element abundances well except N
and Na, but they require significant fine-tuning for each EMP star, with
the appropriate ratio of mixing, fallback, asymmetry and/or jets in the SN
model. The post-formation scenarios include the dredge-up of elements
during the star's giant branch phase, accretion from the interstellar
medium or a companion star, and the rotation of the parent star -- all of
these could enhance C, N, O and sometimes Mg given enough time.  The joint
trend of enhanced Mg, Si and O at low [Fe/H] however remains
unexplained. The strong variations between individual EMPs amongst the
elements shown here and those related to the Fe-peak and $r$- and
$s$-process elements imply that any successful scenario may require
considerable tailoring for each star.

Although N is not an element of much relevance to dust, we include it here,
as it is usually a good tracer of C and O, owing to its origin in the CNO
cycle, and hence of the AGB scenario.  We note that we have excluded stars
known or thought to be in binaries, so we do not necessarily expect to see
the signature from AGBs in binaries in Figure 1.  Unfortunately, the data
on N in this range of [Fe/H] is fairly scant, and no firm trend can be
stated. As noted by \citet{plezcohen}, EMP stars with enhanced C usually
have enhanced N as well, but the reverse is not always true. This is seen
in Figure 1. It is possible that N depletion can occur through the
formation of ammonia (NH$_3$) ices as mantles on grains in dense clouds,
but these likely are sputtered away immediately.

In contrast, although not shown here, it is worth noting that
for the few stars with measured CNO that have [Fe/H] $\ga -3$, [C/Fe] is
mildly underabundant, whereas [N/Fe] is mildly overabundant, an
anticorrelation noted by \citet{briley} in a study of Galactic
globular clusters. The varied behavior of N/C and the lack of a trend with
[Fe/H] in EMP stars is noted in \citet{plezcohen}, who point out that the
SN origin is attractive owing to the additional enhancement of Na, Mg, and
especially O in these stars. The increasing enhancement of N accompanied by
a decline of C as the star's evolutionary state advances would argue for
self-enrichment from dredge-ups and other processes in the giant branch
phase, but this is not observed.

In summary, the data on EMP stars indicate that Si, Mg, and O, and to a
lesser degree C, are generally enhanced in the iron-poor stars. These
elements form the composition of the dominant dust compounds created in SNe
from metal-free stars of masses $\sim$ 10--260 $M_\odot$ \citep{todini,
nozawaetal03, schneideretal04}, which include graphites, silicates
(enstatites (MgSiO$_3$), forsterites (Mg$_2$SiO$_4$), and SiO$_2$), and
iron-bearing magnetites. We therefore proceed to investigate whether these
elements can be selectively transported in a hot radiation field, decoupled
from the background SN metals in the gas phase.

\section{Creation and Transport of Dust Grains}

In this section, we describe the scenario in which dust is created in
primordial SNe, followed by a summary of the assumptions and equations we
use to solve for the transport of dust grains. 

\subsection{Dust Formation}

As discussed earlier, the duration of metal-free star formation is
thought to be $\sim$ 10--100 Myr. The lower limit of a few--10 Myr has
been derived using semianalytic calculations of halo self-enrichment from
SN ejecta \citep{tvs04}, and from 3D gas hydrodynamic simulations of the
chemodynamical feedback from the first SNe on parsec scales in the ISM of
primordial galaxies \citep{wadavenk}. In both analyses, this corresponds to
the intrahalo timescale for the reincorporation of metals created by the
first SNe into cold starforming gas clumps up to metallicities roughly
corresponding to $Z_{\rm cr}$ in primordial galaxies. The upper limit of
100 Myr corresponds to the inter-halo enrichment timescale over which
metals are transported to and pollute neighboring halos at $z \sim$ 10--20
\citep{tvs04,byh03}.

Such estimates are consistent with the 1-50 Myr timescales needed for the
instability in SN shells to set in for second-generation star formation
\citep{salvaterra04}. Stated another way, this timescale is a requirement
for the model in this work, where the dust generated in the very first SNe,
presumably on timescales of order 2--5 Myr, is exposed to the hot radiation
field from a zero-metallicity stellar cluster which is assumed to be within
the SNR.  We focus on the effects of an individual SN  in a stellar
cluster and assume for simplicity that it occurs at the center of its host
galaxy. A Pop III cluster is likely to form with only a few stars in the
densest part of the galaxy, and the massive stars in the Pop III IMF likely
have not migrated too far from their birthsites at $z \gg 10$.  The
specific location may not matter much for this work as we are interested in
relatively local rather than intergalactic transport.  

A related issue is the spatial overlap and relative centers of the SNR
and the background radiation field.  In the scenario we propose here, the
net radiation pressure within the SNR is important, and we assume that the
radiation field of the cluster is not dominated by sources that are
spherically symmetric external to the SNR (so as to cancel the field in the
SNR interior). The Jeans length in primordial gas is of order 1 pc, a
distance that the radii of SN shells in our calculations exceed early in
the SNR evolution (as shown below). Therefore, we assume approximate
spherical symmetry, and that the Pop III cluster is not strongly off-center
relative to the SNR.


The first SNe are required in this model to occur in halos whose virial
temperature is at least 10$^{3-4}$ K, for two reasons. First, this is a
criterion in the model of triggered star formation that we use here
\citep{salvaterra04}. Second, the SN kinetic energies considered here
(10$^{51}$--10$^{53}$ erg) do not exceed the binding energies of such
galaxies, ensuring that we are dealing with subgalactic phenomena rather
than dust transport to the IGM.  A single $10^{53}$ erg SN can expel
gas efficiently from low-mass minihalos of virial temperatures 100 K
(corresponding masses of $\sim 10^6 M_\odot$); even such small halos can,
however, partly survive a $10^{51}$ erg SN \citep{byh03}.

We assume that the dust from the first SNe is created on timescales of
hundreds of days after the SN event \citep{todini, nozawaetal03,
schneideretal04} and solve for the density and temperature of the SNR as
detailed below.  In young remnants, the dust grains presumably form in
cooling, metal-rich ejecta, which are slowly decelerated by interactions
with gas in the SNR interior and by the reverse shocks.  After the dust
grains form through nucleation of ejecta material, they are subject to
destruction through thermal sputtering by plasma (H, He ions) and by
passage of the reverse shock through the dense ejecta.  We compute the
sputtering within the SNR (Sedov-Taylor) interior, but do not compute the
effects of reverse shocks.  Some dust grains may survive these shocks, if a
sufficient number reside in dense cold clumps. This may be indicated by
observations of the Cas~A SNR \citep{greidanus91,hinesetal04}, although
whether the detected dust belongs to the SNR or to circumstellar material
is controversial \citep{krauseetal04}.

Those dust grains that survive are radiatively driven by the Pop
III radiation field, and transported to varying distances as detailed
below. We do not model the effects of the inhomogeneous density and
radiation field arising from multiple stellar clusters, or the nonuniform
density and velocity structure within the SNR. We assume that the initial
velocities of all the dust grains are of order 100 km s$^{-1}$, consistent
with observations of the Cas A SNR \citep{fesenetal87}. We discuss the
effects of varying the initial velocity of the dust grains in more detail
in the next section.

For metal-free Type II SNe \citep{todini}, magnetites and ACG (graphites) in
general dominate the dust mass in the stellar progenitor range 12--35
$M_\odot$, with an increasing magnetite yield with rising SN explosion
energies for 22--35 $M_\odot$. For the specific case of 22 $M_\odot$, the
typical grain sizes are about 0.001 $\mu$m for magnetites, slightly lower
values for silicates, and about 300 \AA\ for graphites. In contrast, the
dust mass from PISNe is strongly dominated by silicates, with a small
nearly constant yield of ACG and a rapidly rising yield of magnetites with
increasing stellar mass over 140--260 $M_\odot$. In addition, a 149 (250)
$M_\odot$ parent star generates post-SN dust grains of characteristic sizes
of 0.001 (0.01) $\mu$m, 0.01--0.1 (0.001--0.01) $\mu$m, and 0.01--0.1 (0.1)
$\mu$m for magnetites, silicates and graphites. These trends of grain
compound and size with stellar mass from \citet{nozawaetal03} and
\citet{schneideretal04} use predictions of element nucleosynthesis from
current hydrodynamical SN models. These results combined with calculations
of the chemistry, temperature and density evolution of the SN ejecta
determine which and when dust compounds can condense after the SN
explosion, and their subsequent growth. The temperature affects the order
of grain formation which in turn influences their characteristic sizes.  In
the PISNe mass range, graphites condense first (at higher ejecta
temperatures), and therefore tend to be larger in size relative to
silicates and magnetites which condense later.  Thus, not all dust grains
are created equal: there is a strong dependence of grain size and net yield
on the star's mass, a point whose importance to the EMP stellar abundances
is demonstrated below.

\subsection{Transport of Dust Grains}

The equation
of motion of nonrotating spherical grains, of mass $m_d={4 \pi \over 3} \rho_g a^3$ (with density
$\rho_g$ and radius $a$), under the effect of radiation pressure,
gravity and gas drag is given by,
\begin{equation}
m_d {d v \over dt}=F_r+F_g-F_{drag} \,. 
\label{eq:motion}
\end{equation}
Here the force $F_r$ due to radiation pressure for a given source of
luminosity $L$, acting on a grain at a distance $r$, is,
\begin{equation}
F_r={L \over 4 \pi r^2 c} \pi a^2 \bar{Q}_{pr},,
\end{equation}
where $\bar{Q}_{pr}=\bar{Q}_{abs}+Q_{sca}[1-\langle \cos \theta \rangle ]$.
The averaging is done over the spectral energy distribution of the radiation
field. We assume a primordial composition for the gas.

$Q_{abs}$ and $Q_{sca}$ are the absorption and scattering coefficient and
$\langle \cos \theta \rangle$ characterizes the scattering property of the
grain. The values of $Q_{abs}, Q_{sca}$, and $\langle \cos \theta \rangle$
for graphites and silicates are tabulated in \citet{draine85}. The values
of $Q_{sca}$ are typically much smaller than $Q_{abs}$ and for our
calculation we have used $Q_{sca}=0$, which makes the estimate of the
radiation pressure on grains conservative. For the values of $Q_{abs}$ for
silicates, we use the fit provided by \citet{ferrdettmar}, and for
graphites, we use the fits provided by \citet{nathetal99}, for different
ranges of incident photon energy.  The radiation field is assumed to be
Planckian with a radiation temperature $T_\ast$.

We have performed similar calculations for magnetites (Fe$_3$O$_4$). This
compound could be produced in primordial SNe
\citep{nozawaetal03,schneideretal04}, subject to the uncertainties in the
explosion mechanism and mass cut of SNe associated with the first
stars. For this work, the role of magnetites is particularly important to
consider, as the differential transport of graphite and silicate dust
grains relative to those in iron compounds must be accomplished if we wish
to explain the origin of the ultra-iron-poor EMP stars that have highly
enhanced C, O, Mg and Si relative to solar values.  For
magnetites, we have used the appropriate values of the absorption
coefficients calculated from Mie scattering theory (S. Bianchi 2005,
private communication).

The gravitational force on the grain is calculated assuming a NFW 
profile \citep{nfw97}, assuming the source of radiation to 
be at the cluster center, as,

\begin{equation}
F_g=-{GM_{tot}(r) m_d \over r^2} \,,
\end{equation}
where $M_{tot}(r)$ is the total mass inside the radius $r$. Following
\citet{komatsu}, one can define a characteristic radius $r_s=r_{vir}/c$ for an object
of total mass $M$ collapsing at a redshift $z$, and for a concentration
parameter $c$, where,

\begin{equation}
r_{vir}^3={M \over (4 \pi /3) \Delta_c(z) \Omega_m \rho_c(z)} \,
\end{equation}
and $\Delta_c(z)$ is the overdensity and $\rho_c(z)=3H^2/8\pi G$ 
is the critical density
of the universe. The overdensity $\Delta_c(z)\approx (18 \pi^2+82x-39x^2)/
\Omega(z)$, with $x=\Omega(z)-1$ and $\Omega(z)$ as the ratio of mean matter
density to critical density at redshift $z$ \citep{bullock01}. The density
at this characteristic radius $r_s$ is given by,
$\rho_s=[c^3 M/(4 \pi r_{vir}^3 m(c))]$ where $m(c)=ln(1+c)-r/(1+c)$. The
total mass within a radius $r$ can then be written as,

\begin{equation}
M_{tot}(r)=4 \pi \rho_s r_s^3 m(r/r_s) \,.
\end{equation}

To compute the drag force, we first calculate the charge on the grain
as detailed further in this section, and use equation (4) from
\citet{drainesalpeter1}. We refer the reader to this paper for greater
detail on this calculation.  The drag force on the dust grain has two
components: collisional drag caused by the physical collision of grains
with H and He atoms, and plasma drag arising from the long-range Coulomb
forces associated with the grain charge. Additional effects for the grain
charge that arise from the grain's motion \citep{shull78} are relevant only
when the grain's velocity exceeds the thermal velocity of protons, and are
not considered here.

Grains are charged by the photoelectric effect in the presence of the
radiation field, and collisions with electrons and protons. Coulomb drag is
the dominant process in this calculation and strongly effects the motion
and evolution of the grains.  We do not include the effects of magnetic
fields in primordial galaxies in this calculation, in the absence of a
compelling theory as to such a field's structure and magnitude and the role
of galactic amplification processes.  The charging from the photoelectric
effect depends on $Q_{abs}$, the absorption coefficient of the grain (see
above) and the radiation field $J_\nu$. For a thermal gas with electron and
proton densities $n_e, n_p$ and temperature $T$, the photoelectric current
is given by,

\begin{equation}
J_{ph}=\int_{\nu_{min}} ^\infty Q_{abs} (a,\nu) y_\nu {4 \pi J_\nu \over h \nu} \, d\nu \,,
\end{equation}
where $\nu_{min}=(w+eU)/h$ corresponds to the minimum energy for which
electrons can escape from the surface, $w$ is the workfunction, and
$y_\nu$ is the normalized photo-yield. We use the fits for $y_\nu$ provided
by \citet{bakes94}.

Following \citet{drainesalpeter2}, we write the currents due to collisions
with protons, as (with $x=eU/kT$, $U=Z_{\rm gr}e/a$ being the grain
potential, and $Z_{\rm gr}$ the grain charge),
\begin{equation}
J_p\approx n_p \Bigl ({kT \over 2 \pi m_p} \Bigr ) ^{1/2} \,,
\end{equation}
and with electrons, as,
\begin{equation}
J_e=-n_e \Bigl ({kT \over 2 \pi m_e} \Bigr ) ^{1/2} \,.
\end{equation}
These currents are enhanced by Coulomb interactions by a factor $g(x)$.
For electrons, $g_e(x)=\exp (x)$, for $x<0$, and $g_e(x)=1+x$ for $x>0$,
and for protons, $g_p(x)=\exp(-Z_{\rm gr}x)$ for $x>0$, and $g_p(x)=1-Z_{\rm gr}x$ for $x<0$
\citep{spitzer}.

The charging time scales are much smaller than the transport time scale of
the grains in our calculations, and therefore the grain charge can be
calculated from the equilibrium condition at each radii : $J_{ph}+J_p=J_e$.

In addition to the equation of motion (equation \ref{eq:motion}), we also
calculate the evolution in grain sizes due to sputtering.  Grains are
considered to be destroyed when they are reduced to $10^{-3}$ of their
original size. For the case of magnetite grains, there are no published
sputtering yields in the literature, and a detailed calculation of this
quantity is beyond the scope of this work.  \citet{bianchi05} derived that
graphite and olivine [(Mg,Fe)$_2$SiO$_4$] grains have similar sputtering
rates as long as the grain velocities are at least 100 km s$^{-1}$, as we
nhave assumed here. The theoretical expectation would be that the values for
magnetite sputtering lie between those for silicates, which have similar
binding energies per atom, and for metallic Fe (Draine 2005, private
communication). The latter may be preferred if oxygen is preferentially
sputtered from magnetite grains.  For these reasons, we approximate the
magnetite sputtering rate as that for pure Fe from \citet{tielensetal}; for
consistency, we use the sputtering rates from these authors' calcuations
for graphites and silicates as well (see also equation 3 of
\citealt{seabshull}).

\section{Results}

We present the main findings of our paper in this section in four
parts in the following order: (1) the role played by a grain's area to mass
ratio in determining its response to a radiation field; (2) solving for the
evolution of the densities and temperatures within primordial SNRs which
will provide the background conditions for grain transport; (3) using the
results from (2) and earlier sections to derive the efficient transport of
graphite and silicate grains relative to magnetites within SNRs, with the
associated implications for the abundance ratios of Fe-poor EMP stars and
the first-stars IMF; and (4) quantifying the impact of varying a few of our
model assumptions on the results in this work. At the end of each of the
first three subsections, we present a brief summary of its findings in the
broader context of the paper.

\subsection{Radiation Pressure on Dust Grains}

The dynamics of dust grains are determined by a complex interplay between
gas density, temperature, and grain properties like size, mass, density,
and charging rate.  The relevant considerations for this problem are: (1)
the ratio of the grain's effective area to mass, which is important for
grain dynamics under radiation pressure and gravity; and (2) the relative
values of the outward force due to radiation and gas drag, another
important input to the grain's motion.

The effective area, also frequently referred to as the absorption cross
section, is determined by the geometrical area ($\pi a^2$) and the
absorption coefficient $Q_{abs}$. We plot the ratio of the effective area
to the mass of the grains (in units of cm$^2$ g$^{-1}$) as functions of the
wavelength of the incident radiation in Figure 2 for graphites, silicates,
and magnetite grains.  For each species, we show two cases with grain radii
$a=0.01 \, \mu $m and $a=0.01 \, \mu $m. We should note that for a
Planckian radiation field with effective temperature $T_{\rm eff}=5\times
10^4$ K, the peak of the spectrum is at $\lambda \sim hc/(2.82 \, kT_{\rm
eff}) \sim 0.1 \, \mu $m. Such a hot radiation field is easily achieved in
the atmospheres of metal-free stars of masses exceeding about ten solar
masses \citep{tsv, bkl}. The curves show that smaller grains in general
have larger ratios of area to mass, and graphites have the largest ratio,
followed by silicates and then magnetites, a direct consequence of their
respective grain densities of 2.2, 3.2, and 5.2 g cm$^{-3}$.  \\

\centerline{\epsfxsize=1.0\hsize{\epsfbox{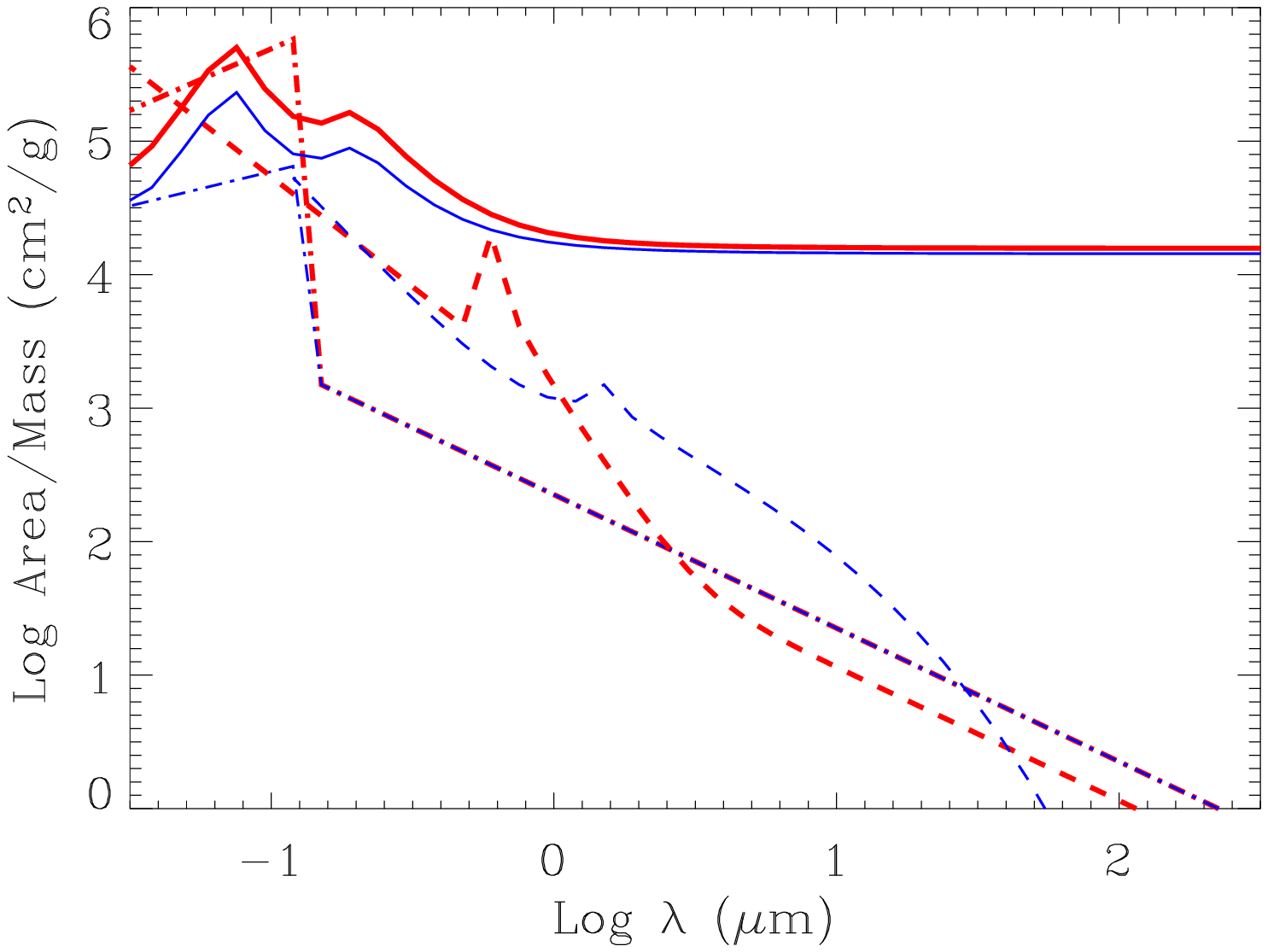}}}

\vspace{0.1in}
\figcaption{The ratio of effective area to grain mass is plotted as a
function of wavelength of the incident radiation, for graphites (solid
lines), silicates (dotted-dashed), and magnetite (dashed) grains.  In each
case, thick and thin curves denote grains of size $a=0.001 \, \mu $m and
$a=0.01 \, \mu $m.  }
\vspace{0.2in}

We therefore anticipate, in the context of SNRs resulting from Pop III
SNe in primordial galaxies, that graphite grains will be affected by
radiation pressure more than silicate and magnetite grains, and small
grains will experience greater radiation pressure than large grains.

\subsection{SNR Evolution}

In order to calculate the transport of dust grains with the equations
described in the previous section, we need to assume approximate background
gas temperatures and number densities within the SNR.  Although we have not
modeled the detailed density and velocity structure in this region, we
assume parameter values that encapsulate the range of physical conditions
expected within the SNR in the SN-induced star formation scenario. We show
this in Figure 3, where the evolution of the density and temperature with
interior radius in the SNR (and not the SN shell radius) is plotted for
times up to 10 Myr. We use Sedov-Taylor self-similar solutions and the
analytic approximations of \citet{kahn75} and \citet{petruk00} for the
temperature and density profiles in Figure 3, assuming a typical HN/PISN
explosion energy $E_{\rm SN} = 10^{52}$ erg and an ambient gas density of
10 cm$^{-3}$.  These parameters represent the typical conditions under
which the fragmentation instability can occur in SN shells in primordial
galaxies \citep{salvaterra04}, although densities of 10 cm$^{-3}$ may only
be achieved in the cold star-forming cores of such galaxies. \\

\centerline{\epsfxsize=1.0\hsize{\epsfbox{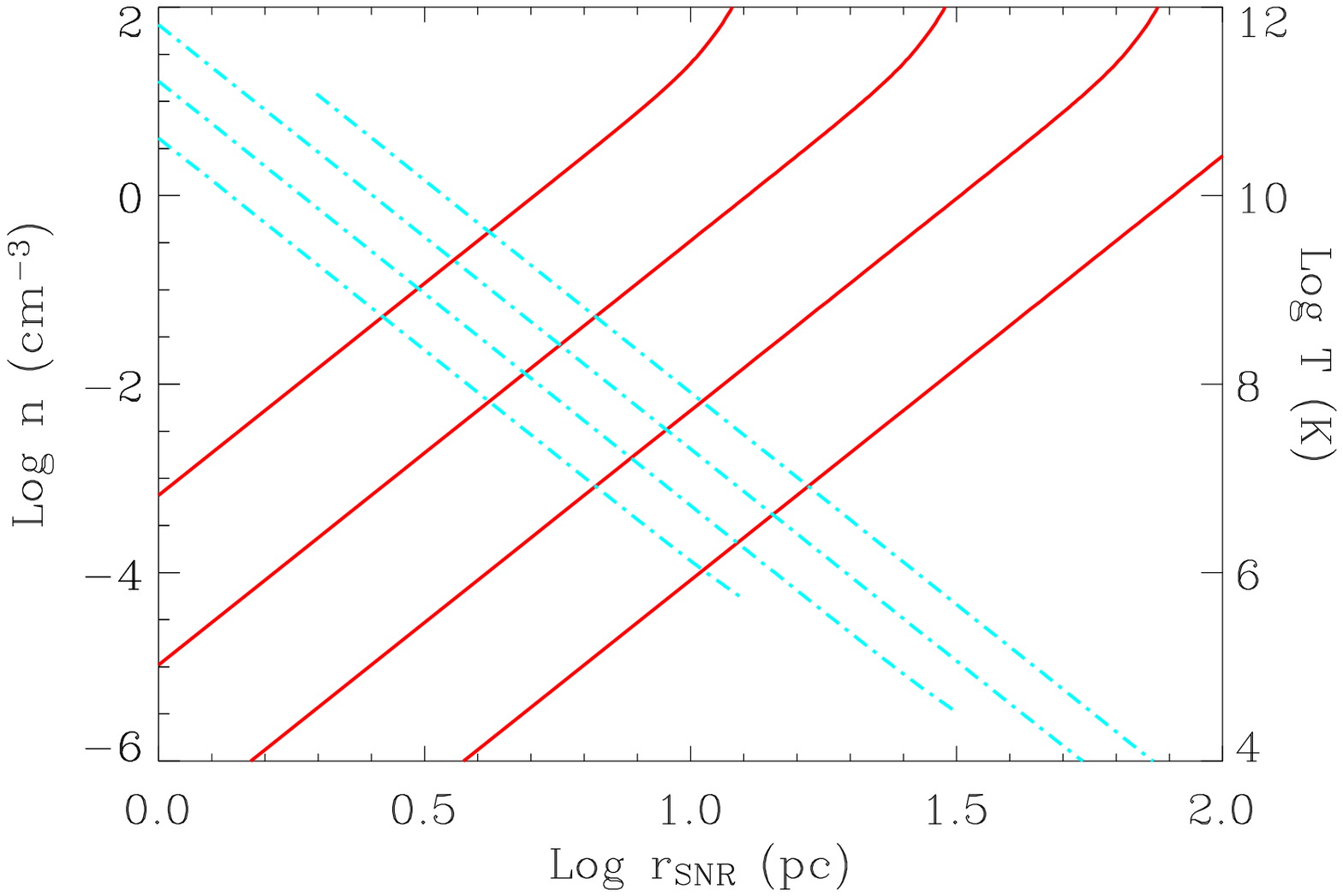}}}

\vspace{0.1in} \figcaption{The evolution of the particle density (solid
  lines; left $y$-axis) and temperature (dashed-dotted lines; right
  $y$-axis) in the SNR interior with radius in pc, for an ambient density
  of 10 cm$^{-3}$ and $E_{\rm SN} = 10^{52}$ erg, and for an assumed
  Sedov-Taylor self-similar solution.  Within each set of lines, the curves
  from left to right are for $t=10^4,10^5, 10^6$ and $10^7$ yr; these would
  move to the right for lower ambient densities (see text).}
\vspace{0.2in}

The timescales on which SN shells form  can be estimated from the
equations describing the Sedov-Taylor phase of SNRs \citep{shullsilk}. 
For the above parameters, the shell radius, velocity, and
post-shock temperature in the shell scale with $E_{\rm 52}$ (SN energies of 10$^{52}$
erg),  ambient density $n_{10}$ (in units of 10 cm$^{-3}$) and time $t_4$ (in units of $10^4$
yr) as:

\begin{equation}
R_s =  (12.7 \; {\rm pc}) \; E_{52}^{0.2} n_{10}^{-0.2} t_4^{0.4}
\end{equation}
\begin{equation}
V_s = (439 \; {\rm km \; s^{-1}}) \; E_{52}^{0.2} n_{10}^{-0.2} t_4^{-0.6}  
\end{equation}
\begin{equation}
T_s = (2.58 \times 10^6 \; {\rm K}) \; E_{52}^{0.4} n_{10}^{-0.4} t_4^{-1.2}  
\end{equation}

Applying the cooling criterion for the onset of
radiative shell-formation with free-free cooling from H$^+$ and He$^{+2}$
ions leads to shell formation at times of:

\begin{equation}
t_{\rm shell} \sim (2.1 \times 10^4 \; {\rm yr}) \; E_{52}^{1/8} n_{10}^{-3/4}  
\end{equation}

Thus, the SN shell has begun to form at about $10^4$ yr, and the
temperature has dropped sufficiently to lower sputtering rates
within the shell. Figure 3 shows that, under these conditions and for the
SN energies considered here, the typical size of SN shells is on the order
of tens to at most hundreds of pc over timescales of 10 Myr. We also
investigated cases with ambient densities of 1 cm$^{-3}$, where the lines
in Figure 3 simply scale towards the right. At a given time and SN shell
radius, the density within the SNR is almost two orders of magnitude lower
in comparison to the 10 cm$^{-3}$ case.

Figure 3 also reveals that at early times ($\la 10^5$ yr) when the SN shell
has advanced to about a few to 10 pc, typical densities and temperatures in
the SNR interior are respectively on the order of 0.1 cm$^{-3}$ and a few
times $10^8$ K. Since the dust that survives post-SN processes such as the
reverse shock is likely to be found in cooler gas, we conservatively assume
ambient temperatures of $3 \times 10^7$ K (the value at about 6 pc at
$10^4$ yr), and an ambient density of 0.1 cm$^{-3}$ within the SNR.

In summary, for a typical Pop III SN with $E_{\rm SN} = 10^{52}$ erg, a SN
shell begins to form in primordial galaxies at $\sim 10^4$ yr after the
initial explosion, and reaches distances of $\sim$ 10--100 pc in 10
Myr. The SNR interior remains hot on these timescales, and we assume
background gas conditions of $3 \times 10^7$ K and 0.1 cm$^{-3}$ in solving
for dust grain transport within the SNR.

\subsection{Grain Dynamics}

We use the results of \citet{schneideretal04} and \citet{todini} for the
formation of dust grains in SNe in primeval galaxies in order to assume the
sizes of grains of different composition in our calculations. They found,
as discussed earlier in this work, that for SNe with a progenitor mass of
149 $M_{\odot}$, the size distribution of magnetite grains peaked at $\sim
0.001 \, \mu$m, and those of graphites and silicates at $\sim 0.04 \,
\mu$m.  For stellar masses of 22 $M_\odot$ and 250 $M_\odot$, the
characteristic grain sizes for magnetites/silicates/graphites are
respectively of order 0.001/0.001/0.03 $\mu$m and 0.01/0.01/0.2 $\mu$m.
These three masses span the range of SN progenitors considered here.  We
refer the reader to these papers and to \citet{nozawaetal03} for the detailed
grain size distribution for various stellar masses. \\

\centerline{\epsfxsize=1.0\hsize{\epsfbox{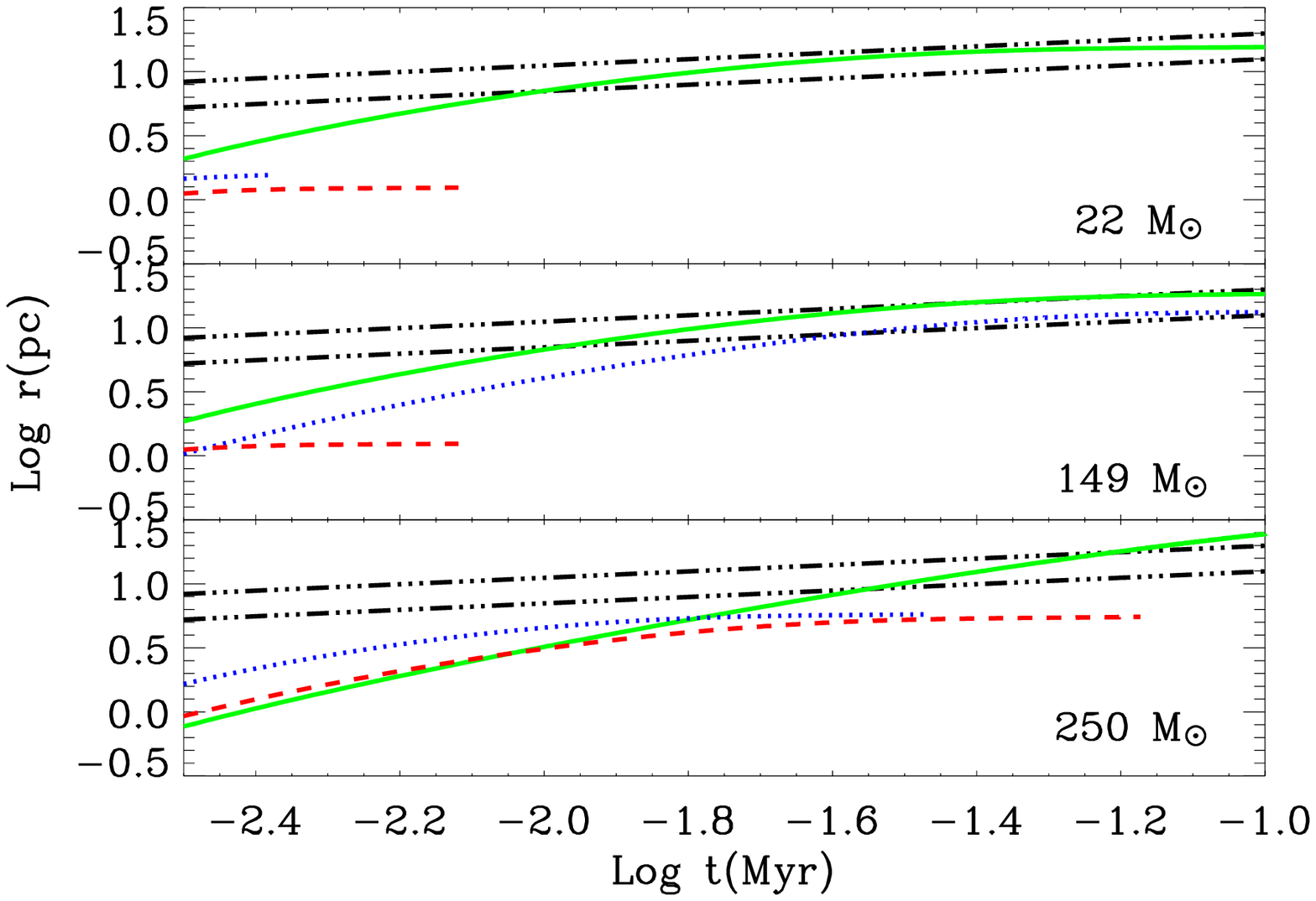}}}

\vspace{0.1in} 
\figcaption{The positions of dust grains within SNRs in primordial galaxies
are shown as a function of time as they are radiatively transported outward
from the center of the SNR. Top, middle and bottom panels represent stellar
masses of 22 $M_\odot$, 149 $M_\odot$, and 250 $M_\odot$, spanning the
range of SN progenitors considered in this work. Solid, dotted and dashed
curves refer to graphites, silicates and magnetite grains.  The initial
grain sizes for the compounds in each panel are detailed in the text.  The
two dashed-dotted lines in each panel describe the adiabatic evolution of
the SN shell radius, for $E_{\rm SN} = 10^{52}$ erg and and ambient gas
number densities of 1 cm$^{-3}$ (upper line) and 10 cm$^{-3}$ (lower
line). See text for discussion.}
\vspace{0.2in}

We show in Figure 4 the position of dust grains under the combined forces
of radiation, gas drag, gravity, and sputtering, using the above values for
the initial sizes of grains of the relevant composition for the three
representative stellar mass cases. We recall our model assumptions
discussed earlier: that all grains are released at $r=0$ with an initial
speed of $100$ km s$^{-1}$, that the luminosity of the central source is
$L=10^6 \, L_{\odot}$ with a Planckian spectrum of effective temperature
$T_{\rm eff}=5 \times 10^4$ K, and that the temperature and gas density
within the SNR are uniform at $3 \times 10^7$ K and $10^{-1}$ cm$^{-3}$
respectively. We do not use the temperatures and densities from the
self-similar solutions shown in Figure 3 for two reasons. First, strong
departures from a roughly constant product of density and temperature
within the SNR occur only close to the SN shell. Second, the self-similar
profile would indicate extremely high temperatures ($\ga 10^{10}$ K) near
the SNR center, which would destroy all dust grains promptly. Such a
profile is an idealization in the very central regions, and it may not be a
realistic representation of the physical conditions there.

Figure 4 shows the radiative transport of graphites, silicates and
magnetites, as well as two additional curves that represent the adiabatic
evolution of the SN shell for an explosion energy of $10^{52}$ erg and
ambient densities of $1$ cm$^{-3}$ (upper curve) and $10$ cm$^{-3}$ (lower
curve). Clearly, the SN shell advances to smaller distances for higher
ambient densities, which could lead to more effective pile-up of grains.
The SN shell positions are computed using the Sedov-Taylor formalism in the
appendices of \citet{shullsilk} for the first $10^2$ yr, and thereafter by
solving the differential equations (2) and (3) in \citet{salvaterra04}
without the radiative cooling term in eqn. 3. We note that the latter
treatment accounts for the mass swept up by the shell, which is substantial
by $10^4$ yr and considerably slows the forward motion of the shell. Hence
the shell radii values in Figure 4 are significantly smaller than those
from a simple Sedov-Taylor treatment for times exceeding $10^4$ yr, e.g.,
11 versus 19 pc at $10^4$ yr for ambient densities of 1 cm$^{-3}$.

The curves in Figure 4 show that dust grains of sufficient size are driven
by radiation equally effectively irrespective of grain type, and that
graphites, silicates, and magnetites may pile up in the SN shell well
before 1 Myr. Although smaller grains experience radiation pressure more
strongly (Figure 2), they are destroyed on faster timescales owing to
greater relative loss of size.  This is seen clearly in Figure 4, where
graphites consistently reach the SN shell owing to their relatively large
size for all the displayed stellar masses, whereas the small magnetite
grains are destroyed at the high SNR interior temperatures before arriving
at the shell. As the temperature increases, the sputtering rates rise and
compete with the decreasing Coulomb drag.  Sputtering affects graphite and
silicate grains as well, but given their typical grain sizes in the figure,
significantly higher densities and/or temperatures ($\ga 7 \times 10^7$ K)
are required to boost sputtering and prevent shell pile-up of these
grains. This can also be achieved by having the SNe explode into an ISM of
lower ambient densities than we have assumed here, in which case the SN
shell would advance to larger distances, leading to less effective pile-up
of dust grains. Transport through SNR interior temperatures that exceed
$\sim$ few $\times 10^7$ K will destroy grains of similar sizes at roughly
the same radii, and will not cause element segregation.

In the context of EMP stellar abundances, where silicate elements and
carbon were often enhanced with respect to iron, the 149 $M_\odot$ case
appears at first glance to offer the best explanation for reasonable
parameter choices. The 22 $M_\odot$ case favors graphites alone, whereas a
SN from a 250 $M_\odot$ star creates larger grains in general, leading to
the eventual transport of graphites, silicates and magnetites to the SN
shell.  The magnetite grains from 22 $M_\odot$ and 149 $M_\odot$ SNe are
considerably smaller ($\sim$ 0.001 $\mu$m) and are sputtered away before
they can arrive at the SN shell.

To fully reconcile these results with the current EMP data presented in \S
2, there is an additional important factor to consider: the relative masses
in these dust grains formed in primordial SNe. The survival and deposition
of any compound must be considered along with its net yield. As discussed
earlier, magnetites and graphites dominate the dust yield from 12--35
$M_\odot$ SNe, while PISNe from the stellar mass range 140--260 $M_\odot$
create mostly silicates, typically 1--2 orders of magnitude more by final
mass than graphites or magnetites. When these trends are considered with
those in Figure 4, we conclude that the mass range $\sim$ 10--150 $M_\odot$
best explains the enhancement of carbon and silicate elements in Fe-poor
EMP stars, with the lower end of this range being most appropriate for
C-rich Fe-poor EMP stars. High-mass PISNe, as seen in the 250 $M_\odot$
case, would predict a strong enhancement of silicates and iron relative to
carbon, which is not matched by the data.

These results remain approximately true when compared with the predictions
of dust grain sizes and mass yields from other works such as
\citet{nozawaetal03}. These authors predict larger grain sizes and yields
for silicates from a 25 $M_\odot$ star than we have assumed here for a 22
$M_\odot$ star from \citet{todini}, and a mass output of graphites that
depends strongly on the degree of mixing in the SN ejecta. The trends from
\citet{nozawaetal03} would imply that more silicates from metal-free Type
II SNe would reach the SN shell, and that the transport of graphites from
10--260 $M_\odot$ stars would be more variable, depending on the physical
conditions in individual SNe. These outcomes will only strengthen the
connection to the EMP stellar data, where the elements Mg, Si and O are
typically overabundant relative to Fe at low [Fe/H], while this is true for
C in only a fraction of the same EMP stars.

To summarize, current models indicate that primordial SNe create graphite
and silicate grains of larger size and, for some stellar masses, in greater
quantities than magnetites. This directly results in the more efficient
transport of graphites and silicates through the SNR interior to the SN
shells, where they could provide cooling that triggers second-generation
star formation. A comparison with the abundance trends in Fe-poor EMP stars
indicates a first-stars IMF in the mass range 10--150 $M_\odot$, {\it if}
dust transport dominates the element ratios in these stars.

\subsection{Model Assumptions and Variations}

We now address a few remaining issues related to our model assumptions,
starting with our estimate of the magnetite sputtering rate. The
calculation of this quantity has two input parameters -- the density and
sputtering rate of a specific grain species. We recall that currently there
exists no sputtering data for magnetites; the rates for either silicates
(which has a similar binding energy per atom) or for pure metallic Fe can
be used as an approximation. The magnetite grain density may be set to its
measured value of 5.2 g cm$^{-3}$, or to that of pure Fe (7.9 g
cm$^{-3}$). There are therefore four possible combinations of these two
parameters in calculating the sputtering of magnetite (recall that for
magnetites in the above figures, we have assumed grain densities of 5.2 g
cm$^{-3}$ and a pure Fe sputtering rate).

\centerline{\epsfxsize=1.0\hsize{\epsfbox{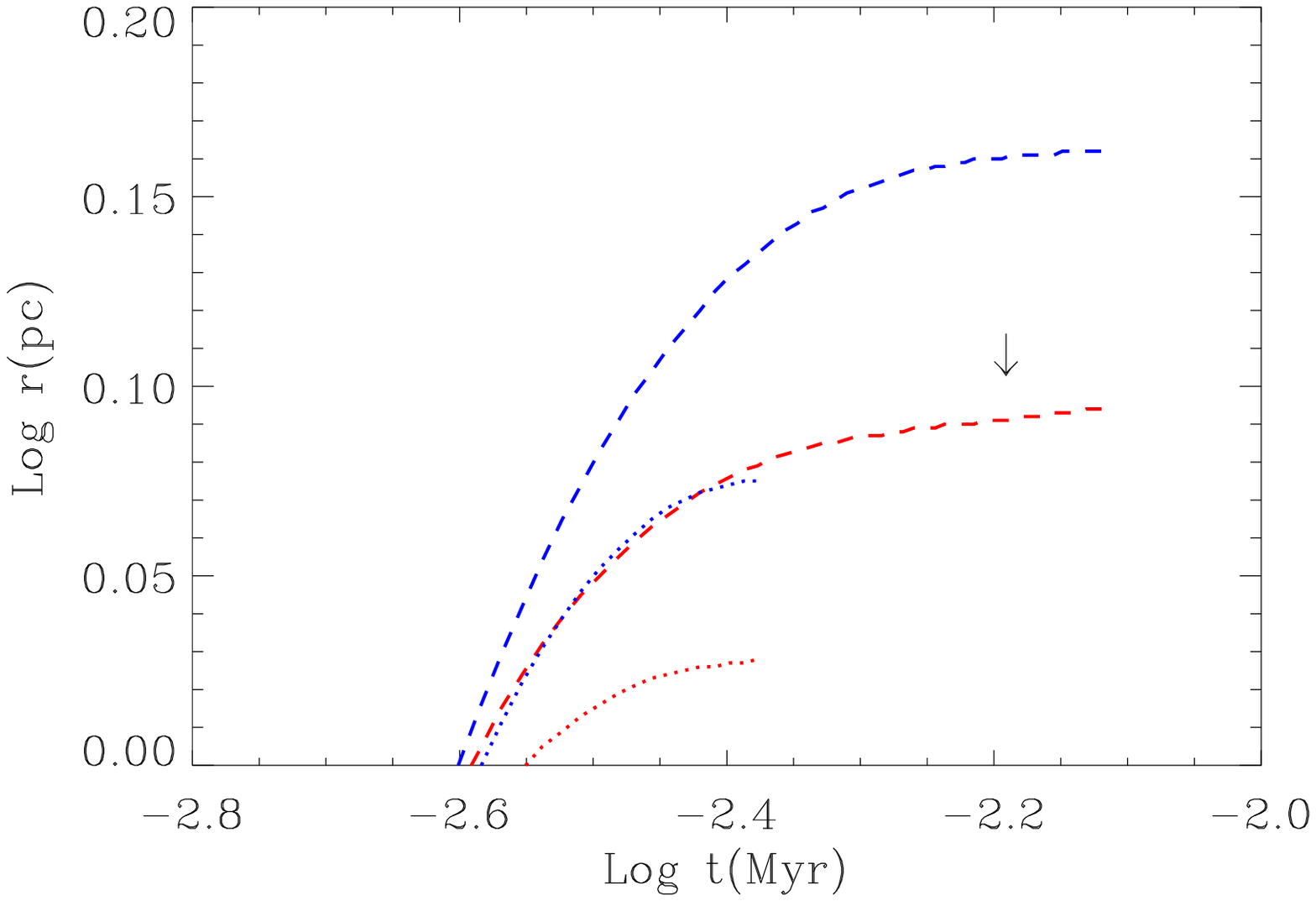}}}

\vspace{0.1in} 
\figcaption{The radiative transport of magnetite grains is shown as a
function of time with varying parameter values for grain density and
sputtering rates. Dotted and dashed lines represent sputtering rates
corresponding to those for silicates and pure Fe respectively; within these
cases, upper and lower curves correspond to densities of 7.9 g cm$^{-3}$  and
5.2 g cm$^{-3}$ (see text). The case assumed throughout the paper is
marked with an arrow and is intermediate to the other cases. The initial
size of all magnetite grains is assumed to be $0.001 \, \mu $m.}
\vspace{0.2in}

We show in Figure 5 the time-dependent position of magnetite grains for
these four parameter combinations for grain sizes of 0.001 $\mu$m.  The
curves which use silicate sputtering rates in general lie below and
terminate earlier than those with pure Fe sputtering rates, as expected
from the higher values of the first quantity. A perhaps less obvious result
in Figure 5 is the larger distance to which the grains with higher density
are transported. This reveals the complex nature of the problem of dust
transport; as we emphasized earlier, individual trends of grain mass, size,
and density do not combine predictably in a given problem. In this case,
the difference between the trajectories of light (less dense) and heavy
(more dense) particles is not simply extrapolated, as they are subjected to
a position-dependent radiation field. One may have expected that heavier
particles always have lower velocities than lighter particles. We
calculated velocities for each of the cases in Figure 5 and find that the
lighter particles do indeed move faster initially, but slow down more
rapidly than do the grains with higher densities.  We attribute this
behavior to the effect of the decreasing values of the driving radiation
field with distance.  Another factor may be the increase of grain charge
with decreasing grain density. This trend appears to hold for the physical
conditions in the problem explored in this paper, where the charge
increases from magnetites to silicates to graphites, leading to the greater
slowing of lighter dust grains from gas drag. This is consistent with the
results of \citet{bianchi05} in the context of dust transport to the
IGM. We emphasize that this anti-correlation between charge and grain
density may be true only in the specific environment considered here, and
may not apply in other astrophysical situations. Our main point which we
wish to demonstrate through Figure 5 is that the curve corresponding to
grain density 5.2 g cm$^{-3}$ and a pure Fe sputtering rate lies
intermediate to the other cases in the figure, and justifies our earlier
assumption of these values for magnetite grains.

Last, we have explored variations of the initial velocities of the dust
grains. Lowering this quantity from our assumed value of 100 km s$^{-1}$ to
10 km s$^{-1}$ does not change our results appreciably, because the speed
to which the grains are accelerated by radiation pressure is itself of
order 100 km s$^{-1}$. However, higher initial velocities of order 1000 km
s$^{-1}$ generally result in the grains being transported to greater
radii. Depending on the role played by grain charge and gas drag, 
the grains may driven out of the galaxy (see earlier
references on dust transport to the IGM), and become less relevant for the
problem of seeding metals for second-generation star formation in early
halos.

\section{Conclusions}

We have investigated the radiative transport of dust within SNRs in
primordial galaxies. We find that the differential transport of the primary
dust compounds from the first SNe could be important for the problem of
which metals seed the second generation of stars that form in the cooling
shells of such SNe. The survivors of such second-generation star formation
may be detected as the EMP stars in the Galactic halo, whose metal
abundance ratios provide clues to the chemical environment in which they
formed. Calculations of the motion of dust grains within SNRs under the
combined forces of the hot first-stars radiation field, gas drag, gravity
and sputtering reveal a complex interplay between grain size, charge, mass,
density, and species.  Our model requires that Pop III star formation
can continue for at least a few to 10 Myr in early halos, the lower limit
on the sum of the lifetimes of massive stars and of the timescale for dust
grains to cross the SN shells (typically well before $10^5$ yr). In the
context of this problem, our main findings are:
\begin{enumerate}
\item The role played by radiation pressure in the grain dynamics of
  different species is strongly influenced by the ratio of the grains' area
  to mass, and decreases from graphites to silicates to magnetites. Within
  a given species, small grains are driven more effectively by radiation
  than are large grains.

\item We find that grain charge increases from magnetites to silicates to
graphites, leading to greater Coulomb drag and loss of velocity for
graphites and silicates relative to magnetites.  However, a competing
effect is the faster destruction of small grains, despite the stronger
radiation pressure experienced by grains of decreasing size. The net result
is that the generally larger graphite and silicate grains from the first
SNe are more effectively driven and accumulate in the SN shells more than
small magnetite grains.

\item If a dust-transport scenario in primordial SNRs were to account for
  the segregation of elements observed in C- and silicate-rich Fe-poor EMP
  stars in the Galactic halo, a metal-free stellar IMF spanning $\sim$
  10--150 $M_\odot$ offers the best explanation. 

\item Higher temperatures and densities than those we have considered here
  will lead to increased sputtering and destruction of most grains, and a
  loss of selective grain transport. At lower ambient ISM densities, the SN
  shell would advance further, reducing the possibility of grains of any
  species reaching the SN shell and mixing with star-forming gas.

\end{enumerate}

A detailed prediction of the contribution of dust to the metallicity of
second-generation stars is beyond the scope of this paper, and is likely
best studied through numerical simulations of primordial star-forming sites
such as those cited earlier in this work. Clearly, dust transport can
enrich such sites in elements that constitute the grains reaching the SN
shell. A 200 $M_\odot$ star can produce up to 60 $M_\odot$ of dust
\citep{schneideretal04}; if this is transported completely to the SN shell
at a radius of 10 pc with an ambient density of 10 cm$^{-3}$ and diluted
into the swept up mass of $\sim 10^3 M_\odot$, then supersolar
metallicities of about 3 $Z_\odot$ in the shell can result. This is only an
estimate of the average value. The real metallicity may be much higher or
lower in cold star-forming clumps in the SN shell, depending on the
efficiency of mixing processes and the role played by some of the
transported elements in providing cooling pathways for primordial gas. 
Although it is beyond the scope of this work to fully compute how and when
the transported dust grains get mixed into the SN shell gas, we may
estimate it from the gravitational instability or free-fall timescale. This
quantity is approximately $(G \rho)^{-1/2}$, and equals 10 Myr for hydrogen
number densities of $\sim 10^2$ cm$^{-3}$. Therefore, if there is to be no
significant delay from dust grain mixing prior to the onset of
second-generation star formation, the gas in the SN shell has to exceed
number densities of about a few $\times 10^3$ cm$^{-3}$.

There remain sizeable uncertainties in the modeling of dust transport in
astrophysical environments, such as those detailed above, and factors
related to the creation of dust in primordial SNe such as the role of
reverse shocks. Clearly, some dust does survive in these environments,
as inferred from the recent detection of dust in a $z \sim 6.2$ QSO whose
extinction curve indicates a SN origin \citep{maiolino04}. An additional
point that we have not accounted for is the binarity fraction of the first
stars, which can affect the mass loss from companion stars, and reverse
shock phenomena related to circumstellar material. Future spectroscopic
analyses of Sloan-DSS and Hamburg/ESO survey data on EMP stars and
observations of a variety of Galactic SNRs will help to address these
important issues and ultimately constrain the formation sites and
conditions of cosmological first- and second-generation star formation.

\acknowledgements

The authors thank Simone Bianchi and Bruce Draine for useful correspondence
on the properties of magnetite, and Yuri Shchekinov for helpful
discussions. We thank an anonymous referee and S. Bianchi for useful
comments on the manuscript. A.~V. gratefully acknowledges the support of
NSF grant AST-0201670 through the NSF Astronomy and Astrophysics
Postdoctoral Fellowship program.  B.~B.~N.  thanks the Fellows of JILA for
their hospitality. J.~M.~S. acknowledges support at the Colorado
astrophysical theory program from NASA LTSA grant NAG5-7262 and NSF grant
AST 02-06042.

\end{document}